\documentstyle[twoside,fleqn,espcrc2]{article}

\newcommand{\AmS}{{\protect\the\textfont2
  A\kern-.1667em\lower.5ex\hbox{M}\kern-.125emS}}

 \def \gta
{\mathrel{\vcenter {\hbox{$>$}\nointerlineskip\hbox{$\sim$}}}}
\title{Supergravity models of Quintessence and Supersymmetry breaking}

\author{Francesca Rosati\address{Centre for Theoretical Physics, 
        University of Sussex, \\
        Falmer, Brighton BN1 9QJ, United Kingdom}%
        \thanks{Present address: Dipartimento di Fisica `Galileo Galilei',
        Universit\`a di Padova, via Marzolo 8, 35131 Padova, Italy. 
        E-mail: francesca.rosati@pd.infn.it}}
       
\begin{document}

\begin{abstract}
The issue of Supersymmetry breaking in the context of Supergravity models of
Quintessence is discussed.
\end{abstract}

\maketitle

\section{INTRODUCTION}

Quintessence cosmology \cite{quint} is probably  one of the best candidates to
fit the most recent data, those that point in the direction of an accelerating
universe  during the present epoch \cite{data}.
Essentially it consists of a scalar field rolling down a runaway potential and
providing the vacuum energy required to accelerate the universe today.
Most
quintessence models make use of the idea of `tracker' fields, i.e scalar
fields rolling down a potential which admits attractor solutions
\cite{scalcosmo,track,scale} thus
relieving the problem of fine--tuning in the initial conditions in this
cosmology. A successfull example is given by inverse power law potentials: 
$V \sim Q^{-\alpha}$, where $Q$ is the Quintessence scalar and $\alpha$ is a
positive number \cite{track}.

Luckily enough, scalar potentials of that kind are found in Supersymmetric
gauge
theories, which exibit a superpotential of the form 
$W =\Lambda^{3+\alpha} Q^{-\alpha}$ ($\Lambda$ is a mass scale),
and have been shown to provide a viable particle physics candidate for
the quintessence scalar \cite{smodel}.
In general, however, quintessence models predict that the VEV of the scalar $Q$
at the present epoch is of order of the Planck mass. If this is the case, we
should take into account Supergravity (SUGRA) corrections to these models.

In this talk I will discuss the issues connected with the construction of 
Supergravity models of quintessence and also address the problem of
Supersymmetry breaking in this context.
In particular, I will report work done in collaboration
with E.~J.~Copeland and N.~J.~Nunes, which is published in ref. \cite{paper}.

\section{SUPERGRAVITY MODELS}

The F-term of the scalar potential in a general SUGRA theory with $n$ $Q_i$
fields is given by the following expression:
\begin{eqnarray}
& & \!\!\!\!\!\!\!\!\!\!  V(Q_i) = |F|^2 - e^{\kappa^2K} 3\kappa^2|W|^2
\nonumber
\\
& & \!\!\!\!\!\!\!\!\!\!  = e^{\kappa^2K} [( W_i +\kappa^2 W K_i ) K^{j^*i}
( W_j + \kappa^2 W K_j )^* \nonumber \\
& & \!\!\!\!\!\!\!\!\!\! - 3\kappa^2|W|^2]
\label{fterm}
\end{eqnarray}
where $F_i= \partial W /\partial Q_i$, the subscript $i$ indicates the
derivative
with respect to
the $i$-th field, and $\kappa^2 = 8\pi G=8\pi M_{\rm Pl}^{-2}$.

Brax and Martin \cite{brax} discuss the  case of a theory with
superpotential $W=\Lambda^{3+\alpha} Q^{-\alpha}$ and a flat
K\"ahler potential, $K = QQ^*$.
It is straightforward to compute the resulting scalar potential:
\begin{eqnarray}
V(Q) &=& e^{\frac{\kappa^2}{2}Q^2} \frac{\Lambda^{4+\beta}}{Q^{\beta}}
\nonumber \\ 
\!\!\!\!\!\!  &\times& \!\!\!\! \left( \frac{(\beta -2)^2}{4} - (\beta +1)
\frac{\kappa^2}{2}Q^2 +
\frac{\kappa^4}{4}Q^4 \right)
\label{flatpot}
\end{eqnarray}
where $\beta = 2\alpha +2\ $.
The main effect of the supergravity corrections is that the scalar
potential can
now become negative due to the presence of the second term.
This is a serious drawback for the model, which becomes ill defined for
$Q \simeq M_{\rm Pl}$.
They go on to propose a possible solution by imposing the condition
that the expectation value of the superpotential
vanishes, $\langle W \rangle =0$. We then see from equation (\ref{fterm})
that the
negative contribution to the scalar potential disappears, and it takes the
form
\begin{equation}
V(Q) \ =\ \frac{\Lambda^{4+\alpha}}{Q^{\alpha}}\,
e^{\frac{\kappa^2}{2}Q^2} \;\; .
\end{equation}
The condition $\langle W \rangle =0$ is possible to realize,
for example, in a model in which we allow 
matter fields
to be present in addition to the quintessence scalar field \cite{brax}.   
Then, if at least one of the gradients of the superpotential with respect
to the matter fields
is non--zero, the scalar potential will always be positive.

This is not the only possibility, though.
There are two obvious problems with the potential (\ref{flatpot}):
one, as already stated, is the negative term in the general
expression (\ref{fterm}) and the second is the choice of the
K\"ahler metric which makes this term grow with the field's vev, relative
to the other terms in the potential.
As mentioned above, setting $\langle W \rangle =0$ is a tight restriction,
so we
will address the issue by relaxing this constraint but allow for more
general forms of the K\"ahler metric.

Such an approach was recently
adopted in \cite{casas96,binetruy97} as a method of obtaining a minimum
for the dilaton field in string theory.
It had the advantage of relying on only one gaugino condensate and
provided an alternative approach to the phenomenology associated
with `racetrack' models \cite{decarlos93}.
In this
scenario, the K\"ahler potential acquires string inspired non-perturbative
corrections.
A further nice feature of these models is that it is possible to
have a minimum with zero
or small positive cosmological constant \cite{binetruy97,barreiro98},
and moreover it is possible to
establish that the dilaton can be stabilized in such a minimum
in a cosmological setting
\cite{barreiro98a}.

In general, for different choices of the K\"ahler metric, the negative
term in (\ref{fterm})
does not always lead to the disaster of a negative minimum in
the scalar potential.
For a general K\"ahler, we do not know {\it a priori}  the
shape of the potential or the location of the minimum.
In fact, in what follows we will show through
explicit examples that the scalar
potential might always remain positive through a
suitable choice of the K\"ahler metric.   
Moreover, with this approach there is no need to introduce additional
fields
in the model.

Let us now go on to study  SUGRA corrections to inverse power law
quintessence models by choosing more general K\"ahler potentials.
Consider, for example, a theory with superpotential
$W = \Lambda^{3+\alpha}\, \tilde{Q}^{-\alpha}$
and a K\"ahler $K = -\ln (\kappa \tilde{Q} + \kappa \tilde{Q}^*)/\kappa^2
\, $,
the type of term which is present at tree level for both
the dilaton and moduli fields in string theory.
In this case, the resulting scalar
potential, expressed in terms of the canonically normalized field
$Q = (\ln \kappa\tilde{Q})/\sqrt{2} \kappa$, is
\begin{equation}
V(Q) = M^4~e^{-\sqrt{2}\beta \kappa Q}
\end{equation}
where $M^4 = \Lambda^{5+\beta} ~\kappa^{1+\beta}~(\beta^2 -3)/2 $ with
$\beta = 2\alpha +1  >  \sqrt{3}$ to allow for positivity of the potential.
This corresponds to the `scaling' solution discussed in the introduction
and so
cannot lead to a negative equation of state for the field in a matter
dominated regime.

Another example follows as a natural extension of the one just described
and leads to potentials with more than one exponential.
For a superpotential of the form
$W = \Lambda^{3+\alpha}\, \tilde{Q}^{-\alpha} +
\Lambda^{3+\beta}\, \tilde{Q}^{-\beta}$
and the same K\"ahler metric as above, then in terms of the same
canonically normalized
field $Q = (\ln \kappa\tilde{Q})/\sqrt{2}\kappa$, the scalar potential
becomes

\begin{eqnarray}
V(Q) &=& (M_1)^4~ e^{-\sqrt{2}\gamma \kappa\, Q} +
(M_2)^4~ e^{-\sqrt{2}\delta \kappa\,
Q}
\\ \nonumber
&+& (M_3)^4~ e^{- \frac{\gamma + \delta}{\sqrt{2}} \kappa\, Q} \,,
\end{eqnarray}
where $\gamma = 2\alpha + 1$, $\delta = 2\beta+1$ and
\begin{eqnarray}
(M_1)^4 &=& \Lambda^{5+\gamma} ~\kappa^{1+\gamma}~(\gamma^2 -3)/2  \,, \\
\nonumber
(M_2)^4 &=& \Lambda^{5+\delta} ~\kappa^{1+\delta}(\delta^2 -3)/2  \,, \\
\nonumber
(M_3)^4 &=& \Lambda^{5+\frac{\gamma+\delta}{2}} ~\kappa^{1+\frac{\gamma +
\delta}{2}}~(\gamma \delta -3)  \,.
\end{eqnarray}

At first sight this appears to be of the form required in \cite{nelson}
in that it involves
multiple exponential terms. However, closer analysis indicates that
the slopes of the exponentials are not adequate to satisfy the
bounds arising from nucleosynthesis constraints, whilst also providing
a positive cosmological constant type contribution today.

As we mentioned earlier, it is possible to have more general
K\"ahler potentials, and with that in mind we now consider the
original model $W = \Lambda^{3+\alpha}\, \tilde{Q}^{-\alpha}$, but
with a K\"ahler potential which depends on a parameter $\gamma$
\begin{equation}
K = \frac{1}{\kappa^2}~\left[ \ln (\kappa\tilde{Q} +
\kappa\tilde{Q}^*) \right]^\gamma \,,
\hspace{1cm}  \gamma > 1 \;\; .
\end{equation}
In this case, the second derivative of the K\"ahler is
\begin{eqnarray}
K_{\tilde{Q}\tilde{Q}^*} &=& \frac{\gamma (\gamma -1)}{\kappa^2
~(\tilde{Q}+ \tilde{Q}^*)^2} \
[ \ln (\kappa \tilde{Q}+ \kappa \tilde{Q}^*)]^{\gamma -2} \, \nonumber \\
&\times& \left( 1- \frac{\ln (\kappa\tilde{Q}+ \kappa \tilde{Q}^*)}{\gamma
-1}
\right)
\label{kinetic}
\end{eqnarray}
and the canonically normalized field $Q$ can be obtained as a
function of $\tilde{Q}$
by integrating the following expression
\begin{equation}
d Q \ =\ \sqrt{2 \, K_{\tilde{Q}\tilde{Q}^*}}\ d\tilde{Q} \;\; .
\end{equation}
In order to avoid the singularity at
$\tilde{Q}+\tilde{Q}^*=1/\kappa^2$,
when $\ln (\kappa\tilde{Q} + \kappa\tilde{Q}^*)$
passes through zero (see equation (\ref{kinetic})), the only possible
choice
is $\gamma =2$.
We then obtain:
\begin{equation}
K_{\tilde{Q}\tilde{Q}^*} \ =\ \frac{2\,
[1- \ln (\kappa\tilde{Q}+\kappa\tilde{Q}^*)]}
{\kappa^2 (\tilde{Q}+\tilde{Q}^*)^2}
\end{equation}
and as a consequence
\begin{equation}
Q \ =\
-\frac{2}{3\kappa} \, [1-\ln (2\kappa\tilde{Q})]^{3/2} \;\; .
\end{equation}
Implying  that the theory is well defined for
\[
-\infty < \ln (2\kappa\tilde{Q}) < 1
\]
which corresponds to $0< \tilde{Q} <e/2\kappa \, $.

The scalar potential in terms of the canonically normalized field $Q$ reads
\begin{eqnarray}
V &=& M^4 ~ \left[ 2x^2 +(4\alpha -7)x+2(\alpha -1)^2
\right]  ~ \frac{1}{x} \nonumber \\
&~&  \times  \exp [(1-x)^2 -2\alpha (1-x)]
\label{vpot}
\end{eqnarray}
where for notational convenience we have defined the quantities
\begin{equation}
x \  \equiv  \left( -\frac{3}{2}\, \kappa Q \right)^{2/3} \ =\
1-\ln (2 \kappa \tilde{Q})  \,,
\label{x-q}
\end{equation}
\begin{equation}
M^4 = \Lambda^{6+2\alpha}~ \kappa^{2+2\alpha}~ 2^{2\alpha} \, .
\end{equation}
Note that the canonically normalized field $Q$ has a range $-\infty <Q<0$.

We can see from equations (\ref{vpot})--(\ref{x-q})
that the scalar potential behaves like
an exponential for $|Q| \gg 1$ and like an inverse power law for
$|Q| \ll 1$, and thus develops a minimum at a finite value $Q_{\rm m}$.
Note that the potential is always positive for any $\alpha > 1.25$.
Thus, we have found that in this case the supergravity corrections induce a
finite minimum in the potential but do not spoil its positivity.
Note also that the field's value in the minimum is exactly
in the region where we
expect the supergravity corrections to become relevant.
For example, with $\alpha =5$
we obtain $Q_{\rm m} \simeq -0.02$ (in $8 \pi G =1$ units),  which
corresponds to
$\tilde{Q} \simeq 1.2$.
Imposing that the minimum of the potential equals the critical energy
density
today, we can also estimate the mass scale $\Lambda$, depending on
$\alpha$.
In the case $\alpha = 5$ we have that $V(Q = Q_{\rm m})
\simeq 10^{-47}\rm{GeV}^4$
which corresponds to $\Lambda \simeq 6~10^{10}\, {\rm GeV}$.

\section{SUPERSYMMETRY BREAKING}

If supersymmetry is to be realized in nature, it must be broken at a
mass scale $M_S$ such that $M_S^2 \sim \langle F \rangle \gta (10^{10}
{\rm GeV})^2$ or   $M_S^2 \sim \langle F \rangle \gta (10^{4}
{\rm GeV})^2$ (for gravity and gauge mediated cases respectively),
in order to lift the supersymmetric scalar particle masses 
 above $10^2 {\rm GeV}$. 

  This then requires the superpotential in (\ref{fterm}) to be
$W \sim\langle F \rangle \kappa^{-1} \sim m_{3/2} \kappa^{-2}$ in order to
cancel
the F-term contribution and consequently to give a negligible total vacuum
energy density ($m_{3/2}$ is the gravitino mass).

  From the discussion in the last section,
it is clear that the dynamical
cosmological constant provided by the quintessence potential cannot
be the dominant source of SUSY breaking in the Universe as
$W \simeq \Lambda^{3+\alpha} \kappa^{-\alpha}\sim (10^{-3} {\rm eV})^2
\kappa^{-1} \ll
\langle F \rangle \kappa^{-1}$. Therefore,
we need some additional source of SUSY breaking. If we consider now
the superpotential,
\begin{equation}
W =  \Lambda^{3+\alpha}~Q^{-\alpha} + m_{3/2} \kappa^{-2} \,,
\end{equation}
then one gains a correction to the scalar potential in (\ref{vpot}) of,
\begin{equation}
\delta V \sim \Lambda^{3+\alpha} m_{3/2} \kappa^{-\alpha} +
m_{3/2}^2 \kappa^{-2} \,,
\end{equation}
for  $Q \sim \rm{M_{Pl}}$ today \cite{choi}.

  The first term can in principle be controlled for sufficiently
 large $\alpha$, however,
the constant second term unavoidably leads to a disruption of the
quintessence
potential. This is a very serious problem of all supergravity models
in quintessence. (In order to avoid the SUGRA corrections
problem, Choi proposes a Goldstone-type quintessence model
inspired in heterotic M-Theory \cite{choi}).

The situation gets worse, since, as pointed out in \cite{lyth},
it appears that even if we
imagine that the amount of SUSY breaking that we observe in the universe
today
comes from a hidden sector other than the quintessence one, there will
still be gravitational couplings between the two sectors that rekindle
the original problem.

However, some recent proposals point in a slightly different
direction for solving
this problem. The basic idea is that the traditional approach
to SUSY breaking might
not be the best way to explain the world we live in.
For example, the mass difference between the superpartners could
arise in a 4D world
with {\it unbroken} SUSY through some higher dimensional effects
\cite{witten}. If
this is the case, we would not need to break supersymmetry, and
the quintessence
potential would be preserved.
Another possibility \cite{banks} is that the relation between
the SUSY breaking scale
$M_S$ and the cosmological constant $\langle F \rangle ^2$ is not what is
usually considered.

If $M_S =  \kappa^{-1} [ \langle F \rangle ^2 {\kappa}^{4}]^{\beta}$
and $\beta = 1/8$,
instead of the
usual $1/4$, then the observed cosmological constant would provide
just the right
amount of SUSY breaking. In this case we wouldn't have any dangerous
F-term of order
$\sim \kappa^{-2} M_S^2$ which spoils the quintessence potential.

\end{document}